\documentstyle[graphicx,float]{article}

\begin{document}
\title{The exponential laws for emission and decaying of entangled atoms}
\date{}
\author{Pedro Sancho \\ Centro de L\'aseres Pulsados CLPU \\ Parque Cient\'{\i}fico, 37085 Villamayor, Salamanca, Spain}
\maketitle
\begin{abstract}
The first photon emission and the disentanglement of a pair of
identical bosonic atoms in excited entangled states follow an
exponential law. We extend the theory to distinguishable and
identical fermionic two-atom systems. As a byproduct of
the analysis we determine the symmetries of the fermionic wave
function. We also derive the emission distributions of excited atoms
in product states, which must take into account the presence of
simultaneous detections. Comparing both distributions reveals a
direct manifestation of the modifications induced by entanglement on
the atomic emission properties.
\end{abstract}

\section{Introduction}

Entanglement plays an important role in the description of
light-matter interactions in multi-particle systems. Several authors
have analyzed different aspects of the problem (\cite{fri,fic,dow} and
references therein), in particular, atomic absorption and emission
\cite{yoE}. Recently, two experiments have studied possible effects
of entanglement in the emission distribution of excited hydrogen
atoms generated in the photodissociation of $H_2$ molecules
\cite{jap,bel}. An interpretation of the experiments has been
presented in \cite{yo}. A fundamental element of this interpretation
is the validity of exponential laws for the first photon emission
and for the decay of the atomic pair to a non-entangled state.

In this paper we explore exponential laws in pairs of identical
fermionic atoms and distinguishable ones. Following the strategy
used for bosons, the analytical form of the decaying parameter can be
evaluated by invoking the compatibility between the emission
distributions with and without temporal ordering. In addition, in
\cite{yo} it was shown that the explicit form of the decaying rate
can be explained by a combination of three effects: entanglement,
exchange effects (the photodissociated atoms are identical bosons)
and disentanglement after the first spontaneous emission (the
spontaneous emission by an atom leaves the multi-particle system in
a non-entangled state \cite{Eb,Ya}). A similar argumentation can be
done here but, because of its similarity with the original one, we
only sketch it in the Appendix. The only novel element of this
argumentation is that, in the case of identical fermions, it
determines the symmetries of the wave function previous to
antisymmetrization.

Later, we consider the emissions by pairs of distinguishable atoms
in excited product states. The evaluation of the distribution demands a
careful consideration of the aspects related to the existence of
simultaneous detections. The distributions for entangled and product
states are different, providing a potential method to verify the
dependence of the emission properties on entanglement. The main
advantage of this approach is that, at variance with the
arrangements \cite{jap,bel}, this dependence appears independently
of exchange effects. We can test without any interfering effect the influence of entanglement.

In the final part of the paper we discuss some conceptual aspects of the problem, in particular, the discontinuity between distributions associated with entangled and product states, the qualitative rather than quantitative influence
of entanglement on emission, the characterization of multi-atom
systems as single emitters, and the differences between exponential
decaying laws and disentanglement in finite times \cite{Eb,Ya}.

\section{Emission in entangled states}

In the case of pairs of excited identical bosonic atoms, the first
photon emission obeys an exponential law \cite{yo}. As the first
emission leads to the disentanglement of the atoms, the temporal
variation of the number of entangled atoms must be expressed via the
standard decaying law:
\begin{equation}
\frac{dn_e}{dt}=-\Gamma _f n_e
\end{equation}
with $n_e$ as the number of entangled pairs and $\Gamma _f$ as the rate of
first emissions. The trivial solution of the equation is $n_e(t)=n_0
\exp (-\Gamma _f t)$, where $n_0$ is the initial number of entangled
pairs.

From the experimental point of view the interesting distribution is
that of first emitted photons, much more simple to measure than that
of entangled pairs. It is given by $N_f(t)=n_0(1-\exp (-\Gamma _f
t))$, which can be derived from $N_f(t)+n_e(t)=n_0$.

We shall explore the consequences of assuming that the decaying law
also holds for pairs of distinguishable atoms. Thus, we assume that
the two above equations for $n_e$ and $N_f$ remain valid when we
consider a pair of distinguishable atoms denoted by $A$ and $B$.

We know the form of the emission distribution for each type of atom.
If we do not compare the emission times of the two atoms, we have a
single-atom process. Only when we compare these times we are dealing
with multi-atom phenomena and a dependence on entanglement can be
present \cite{yo}. Then the distribution of each type of photon is
that associated with an ensemble of single atoms (or pairs of atoms
in product states). The photon distribution is
\begin{equation}
N_i(t) = n_0 (1-e^{-\Gamma _i t})
\label{eq:Nii}
\end{equation}
with $i=A,B$ the two types of photons (denoted by the same symbols of
the emitting atoms), and $\Gamma _i$ the emission rate of atoms of
type $i$.

From these distributions we can obtain the second photon emission
$N_s(t)=N_A(t)+N_B(t)-N_f(t)$. However, in this paper we are not
interested in this distribution.

Next, we must evaluate the rate of first emissions. This can be done by
expressing $N_i$ in terms of $\Gamma _f$. We denote by $n_i$ the
number of excited atoms of type $i$ in product states, that is, the
number of atoms whose companion in the pair (of type $j$, $j \neq
i$) has already emitted but it is yet in the excited state (the
second emission of the pair will be of type $i$). We have
\begin{equation}
\frac{dN_i}{dt}=\gamma _i n_e + \Gamma _i n_i
\label{eq:Ni}
\end{equation}
This expression shows that there are two contributions to the
emission of photons of type $i$. The first one, $\gamma _i n_e $, is
associated with the first emissions. It represents the channel where
the first emissions are of the type $i$, denoting $\gamma _i$ the
rate of first emissions in this channel. The second contribution
refers to non-entangled excited atoms of type $i$ (atoms initially
belonging to entangled pairs where the first emission has been of
the type $j \neq i$) that emit (is the second emission of the pair).
As the atoms are disentangled they emit with the single-atom rate
$\Gamma _i$.

The variable $n_i$ is an additional one in the problem, obeying the equation
\begin{equation}
\frac{dn_i}{dt}=\gamma _j n_e - \Gamma _i n_i     \; \; \; i \neq j
\end{equation}
The number of non-entangled excited $i$ atoms increases by first
emissions in the channel $j$ (first emitted photons of type $j$) and
decreases by second emissions of type $i$ leaving the atom in the
ground state.

This equation can be easily solved. With the initial condition
$n_i(0)=0$ (initially all the atoms are entangled) we obtain
\begin{equation}
n_i(t)=\frac{n_0\gamma _j}{\Gamma _i - \Gamma _f}(e^{-\Gamma _f t}-e^{-\Gamma _i t})
\end{equation}

Using this expression we can rewrite Eq. (\ref{eq:Ni}) as
\begin{equation}
\frac{dN_i}{dt}=n_0\gamma _i e^{-\Gamma _f t} + \Gamma _i \frac{n_0\gamma _j}{\Gamma _i - \Gamma _f}(e^{-\Gamma _f t}-e^{-\Gamma _i t})
\end{equation}
On the other hand, from Eq. (\ref{eq:Nii}), the temporal variation of $N_i$ can be also expressed as
\begin{equation}
\frac{dN_i}{dt}=n_0\Gamma _i e^{-\Gamma _i t}
\end{equation}
Thus, we have obtained two expressions for the temporal variation of
emitted photons of each type. The first one uses the temporal
ordering of the emissions, whereas the second one is independent of
it. The compatibility of both expressions demands them to be equal.
The equality conditions leads to the four relations ($i,j=A,B$)
\begin{equation}
\gamma _j=\Gamma _f - \Gamma _i
\end{equation}
and
\begin{equation}
\gamma _i = \frac{\Gamma _i \gamma _j}{\Gamma _f - \Gamma _i}
\end{equation}
It is immediate to obtain the solution:
\begin{equation}
\gamma _i = \Gamma _i
\end{equation}
and
\begin{equation}
\Gamma _f = \Gamma _A + \Gamma _B
\end{equation}
The emission rate of the first photon by the entangled system is the
sum of the emission rates of the two atoms. This result agrees with
the study  of systems interacting simultaneously with two different
weak noises, where the resulting decay rate is the sum of the
separate rates \cite{Ya}.

The above analysis is also valid for identical atoms. When the two
atoms are identical the emission rate doubles that of the single
atom, recovering the result presented in \cite{yo} for bosons. For
fermions we obtain the same decaying rate, $\Gamma _f^F=2\Gamma _i$.

In \cite{yo} the analytical form of the emission rate was explained
by an explicit calculation based on the matrix elements associated
with the process. Here, we could follow a similar procedure.
However, it would be almost identical and we shall not repeat it. We
only present in the Appendix the distinctive features of the
derivation for fermions and distinguishable atoms.

\section{Emission in product states}

In order to compare with the entangled case, we derive in this
section the photon emission distributions of atoms in product
states. The physical information we have about the physical system
is the exponential form of the single atom emissions and the
statistical independence of the two processes. From these facts we
can deduce the first emission distribution.

We consider the probability of only one emission in a temporal
interval $\tau $ centered around $t$, $P(t,\tau)$. The
temporal interval is dictated by the experimental window of
detection. It can vary from experiment to experiment. As signaled
before, a distinctive characteristic of non-entangled emitters is
that the emissions are statistically independent. An immediate
consequence of this independence is that we can have simultaneous
emissions. For these events we cannot define a first emitted photon
and we must discard these pairs. We restrict our considerations to
the post-selected set with only one emission in each interval. The
unnormalized one emission probability is
\begin{eqnarray}
P^{un}(t,\tau)= P_A(t,\tau)\bar{P}_B(t,\tau) +
\bar{P}_A(t,\tau)P_B(t,\tau) = P_A(t,\tau)(1-P_B(t,\tau)) +
\nonumber \\ (1-P_A(t,\tau))P_B(t,\tau)=  P_A(t,\tau)+P_B(t,\tau) -2
P_A(t,\tau)P_B(t,\tau)
\end{eqnarray}
where $P_i(t,\tau)$ is the probability of emission by the atom
$i=A,B$ in that interval and $\bar{P}_i(t,\tau)$ is the probability
of not emission in the same interval. The superscript $un$ denotes
the unnormalized character of the distribution. The total
probability of one emission is the probability of emission by atom
$A$ and not emission by $B$ plus the probability of the same process
with the roles of the two atoms interchanged. As the emissions by
the two atoms are statistically independent, each one of the two
terms is the product of the events probabilities.

The probabilities can be evaluated from the relation
\begin{equation}
P_i(t,\tau)=\frac{N_i(t+\tau /2)-N_i(t- \tau /2)}{n_0} \approx \frac{\tau }{n_0} \frac{dN_i}{dt} = \tau \Gamma _i e^{-\Gamma _i t}
\label{eq:tre}
\end{equation}
that represents the ratio between the number of photons emitted in
the interval and the total number of emitted photons (of type $i$),
which can be approximated using Taylor's rule.

We must normalize the probability distribution to unity, $\int
_0^{\infty} P(t,\tau)/\tau dt=1$,  because we have assumed that the
simultaneous detections have been removed. A trivial integration
gives
\begin{equation}
\int _0^{\infty} \frac{P^{un}(t,\tau)}{\tau} dt= 2-  \frac{2\tau
\Gamma _A \Gamma _B}{\Gamma _A + \Gamma _B}=\alpha ^{-1}
\label{eq:cat}
\end{equation}
Finally, the one-emission probability becomes
\begin{equation}
P(t,\tau)=\alpha  \tau ( \Gamma _A e^{-\Gamma _A t} + \Gamma _B
e^{-\Gamma _B t} - 2\tau \Gamma _A \Gamma _Be^{-(\Gamma _A + \Gamma
_B) t} )
\end{equation}
The one-emission probability equals the first emission probability.
From it we can easily obtain, with an argument similar to that used
in Eq. (\ref{eq:tre}), the distribution of first photon emissions.
From $P/\tau \approx (dN_f^p/dt)/n_0$ and imposing the initial
condition $N_f^p(0)=0$ we have
\begin{equation}
N_f^p(t)=n_0 \left( 1-\alpha e^{-\Gamma _A t} - \alpha  e^{-\Gamma _B t}+ \alpha \frac{2\tau \Gamma _A \Gamma _B}{\Gamma _A + \Gamma _B} e^{-(\Gamma _A + \Gamma _B) t} \right)
\end{equation}
with the superscript $p$ denoting that now we are dealing with atoms in product states.

In the limit of long times we have $N_f^p(\infty)=n_0$. Note that we
obtain $n_0$ photon detections because we restrict our
considerations to the post-selected ensemble. This is equivalent to
have the total probability of only one detection equal to one. If we
had worked in the total ensemble including the simultaneous
detections we would have obtained a number of first detections
smaller than the initial number of excited atoms.

The first photon distribution for product states is not an
exponential function but the sum of three different ones. In addition to
the temporal dependence typical of single atoms, $\Gamma _A t$ and
$\Gamma _B t$, we also have a dependence on the sum of both rates,
$(\Gamma _A +\Gamma _B )t$. On the other hand, the distribution is
also a function of the experimental temporal window $\tau$. Using
different windows leads to different numbers of removed coincidence
detections.
\begin{figure}[H]
\center
\includegraphics[width=7cm,height=7cm]{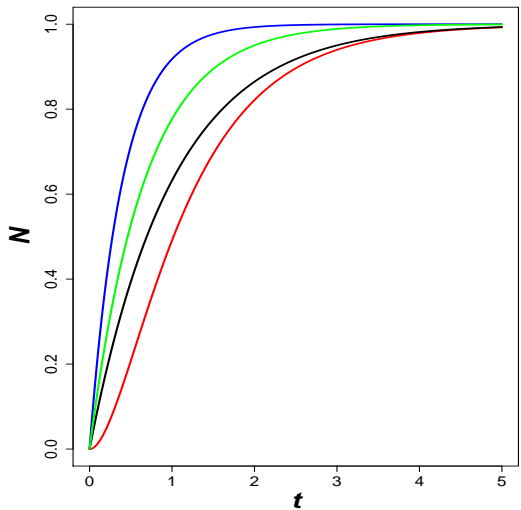}
\caption{Temporal dependence of the normalized photon emission
distributions. The blue, red, black and green curves correspond
respectively to $N_f/n_0$, $N_f^p/n_0$, $N_A/n_0$ and $N_B/n_0$. We
represent the case $\Gamma _A=1$, $\Gamma _B=1.5$ and $\tau = 5/6 $.
The distributions are dimensionless and the time is in units of
$\Gamma _A^{-1}$ .}
\end{figure}
Now, we can compare the first distributions for entangled (with the
assumption of an exponential form) and product states. We present
this graphical comparison in Figure 1. Typical values of atomic
emission rates, $\Gamma $, are of the order of $10^9 \, s^{-1}$
(approximate value for the hydrogen atom). On the other hand, the
temporal windows used in photon detection experiments range usually
between $10^{-8} \, s$ and $10^{-10} \, s$. Note that our method is
only valid for temporal windows obeying $\tau \Gamma _A \Gamma _B <
\Gamma _A + \Gamma _B$ (Eq. \ref{eq:cat}).

We see that the first photon distributions for entangled and product
states of excited distinguishable atoms differ. This result could be
used (provided the experiments verify the exponential form for the
entangled case) to directly test the role of entanglement in
spontaneous emission without the interfering contribution of the
identity effects present in the case of identical particles.

\section{Discussion}

The ideas discussed in this paper are, in principle, testable with
minor modifications of the arrangements in \cite{jap,bel}. In order
to obtain the first-photon distributions for entangled states we
only need to photodissociate molecules composed of identical
fermionic atoms or distinguishable ones. As in \cite{jap,bel}, the
decaying products must be in excited states. This way we can test if
they obey, as in the bosonic case, an exponential law. On the other hand, the
distributions for product states could be obtained in a number of
ways. For instance, we could even use two samples composed each
one of a type of atom and compare the emission times.

The validity of the exponential law does not depend on the initial
degree of entanglement of the two atoms. This result has been
experimentally tested for identical bosonic atoms in \cite{bel} and
has been theoretically derived here for fermionic and
distinguishable ones. We do not have a smooth transition from the
exponential law to a non-exponential one (but a sum of different
exponentials) when the initial degree of entanglement decreases.
There is a discontinuity in the emission distributions for the
transition between entangled and non-entangled states. The type of
decaying law ruling the process is only a function of the presence
or absence of entanglement. We are dealing with a qualitative
property of the system, not with a quantitative one.

The existence of an exponential law for the first emission of a
composed system is also relevant from the fundamental point of view.
In addition to the presence of non-classical correlations, the most
distinctive conceptual characteristic of entangled systems is the
loss of individuality of their components. If a system is entangled
one cannot define the states of the components. Equivalently, one
can ascribe a complete set of properties to the full system, but not
to each one of the components (see \cite{Ghi,fri} for an excellent
discussion of this point in the context of identical particles, but
that can be easily translated to distinguishable ones). Closely
related arguments can be developed in our case. We describe excited
individual atoms as single emitters obeying an exponential law. The
exponential law for the first emission of the two-atom system shows
that the complete system, although composed, behaves like a single
emitter. In contrast, when the two atoms are in product states, the
complete system cannot be viewed as a single emitter. When the atomic
pair is entangled a composed system has a property usually
associated with single entities, reflecting a new aspect of the
subtle relation between the parts and the whole in entangled
systems.

As signaled in the introduction, the experiments of the type
\cite{jap,bel} provide the first demonstration of
disentanglement by spontaneous emission. Similar experiments with
identical fermionic atoms and distinguishable ones would provide a
verification of the phenomenon for these types of atoms. In all the
cases we expect an exponential decaying law for the entanglement.
This behavior contrasts with that described in \cite{Eb} where the
decay occurs, for some values of the problem parameters, in a finite
time. This difference can be explained by the different contexts
associated with both cases. In \cite{Eb} the excited atoms are
placed in two cavities, whereas in our approach the atoms are in
free space (only interacting with the vacuum electromagnetic field).

\section*{Appendix}

{\bf Identical fermions.} For fermions we can follow the
presentation done in \cite{yo} for bosons. The two-atom initial
state (just after the photodissociation in the case of a decaying
molecule) is
\begin{equation}
|\Psi _0^F >=N_0^F(\Psi ({\bf x},{\bf y})- \Psi ({\bf y},{\bf x}))|e>_1|e>_2
\end{equation}
The indices $1$ and $2$ denote the two fermionic atoms and the
spatial variables ${\bf x}$ and ${\bf y}$ their center of mass (CM)
position. $\Psi$ is the CM wave function and $|e>_i, i=1,2$ refers
to the excited internal state of the atoms. The minus sign between
the CM wave functions reflects the antisymmetrization of the full
state. The normalization coefficient is
\begin{equation}
N_0^F =\frac{1}{(2-2Re(<\Psi ({\bf x},{\bf y})|\Psi ({\bf y},{\bf x})>))^{1/2}}
\end{equation}
After the first emission the two-atom state becomes
\begin{equation}
|\Psi _f^F >=\frac{1}{\sqrt 2}(\psi ({\bf x})\phi({\bf y})|g>_1|e>_2
- \psi ({\bf y})\phi({\bf x})|e>_1|g>_2)
\end{equation}
Because of the disentanglement inherent to the emission, the CM wave
functions are product ones \cite{yo}. $|g>_i$ denotes the ground
state of the atoms.

Now, we can obtain the emission rate by evaluating the transition
matrix element $<\Psi _f^F |\hat{U}|\Psi _0^F>$, with $\hat{U}$ as the
evolution operator. This can be done by following step by step the
derivation for bosons in \cite{yo}. It is immediate to see that the
only difference is the sign between the wave functions (and in the
normalization coefficient): it is positive for bosons and negative
for fermions. It is also straightforward to realize that we reach
the relation $\Gamma _f^F =2\Gamma _i$ if the CM wave function
(previous to antisymmetrization) is antisymmetric
\begin{equation}
\Psi ({\bf x},{\bf y})= - \Psi ({\bf y},{\bf x})
\end{equation}
This contrasts with the case of bosons, where the CM wave function
(previous to symmetrization) must be symmetric. Note that for
fermions the wave function must be antisymmetric all the time
between the preparation and the first emission. At the initial time
it is an initial condition. For instance, if the atoms are generated
by dissociation of a molecule it is natural to assume that being the
atoms identical fermions the exchange effects lead to an initial
antisymmetric state. The subsequent free evolution of the atoms
preserves the antisymmetric character of the initial state. In
effect, using the propagator $G$ we can write the wave function at
times $t>0$ as
\begin{equation}
\Psi ({\bf x},{\bf y},t)= \int d^3{\bf x}_* d^3{\bf y}_* G({\bf x},{\bf y},t; {\bf x}_*,{\bf y}_*,0)  \Psi ({\bf x}_*,{\bf y}_*,0)
\end{equation}
For free evolution, the Schr\"odinger equation and, consequently, the
propagator are symmetric. The simultaneous interchange ${\bf x}
\leftrightarrow {\bf y}$ and ${\bf x}_* \leftrightarrow {\bf y}_*$
shows that the wave function at $t$ remains antisymmetric.

We conclude that the relation $\Gamma _f^F =2\Gamma _i$ imposes a
precise symmetry on the wave function (previous to
antisymmetrization). Without that relation the symmetry properties
of the wave function would remain undetermined.

{\bf Distinguishable atoms.} When the atoms are distinguishable, the
initially entangled state $\Phi ({\bf x},{\bf y})|e>_A|e>_B$ can
evolve into the disentangled states $\psi ({\bf x})\phi ({\bf
y})|g>_A|e>_B$ or $\psi _*({\bf x})\phi _*({\bf y})|e>_A|g>_B$. The
two channels for the first emission are distinguishable because the
emitted photons have different frequencies. Then we must add
probabilities instead of probability amplitudes. Following the steps
in \cite{yo} we have that the probability in each channel equals its
emission rate and, finally, we obtain $\Gamma _f =\Gamma _A + \Gamma
_B$.

\end{document}